\documentclass[aps,psfig,epsfig,preprint] {revtex4}
\usepackage{epsfig}
\usepackage{color}
\usepackage{bm}
\usepackage{subfigure}
\usepackage{graphicx}

\begin{document}
\draft
\title{
{\bf Lepton flavor violating $\mu\rightarrow e\gamma$ and $\mu-e$
conversion in unparticle physics }}

\author{Gui-Jun Ding$^{a}$}
\author{ Mu-Lin Yan$^{a,b}$}

\affiliation{\centerline{$^a$Department of Modern
Physics,}\centerline{University of Science and Technology of
China,Hefei, Anhui 230026, China}\centerline{$^b$ Interdisciplinary
Center for Theoretical Study,} \centerline{University of Science and
Technology of China,Hefei, Anhui 230026, China}}

\begin{abstract}
We have studied lepton flavor violation processes $\mu\rightarrow
e\gamma$ and $\mu-e$ conversion in nuclei induced by unparticle.
Both ${\rm Br}(\mu\rightarrow e\gamma)$ and $\mu-e$ conversion rate
${\rm CR}(\mu-e,{\rm Nuclei})$ strongly depend on the scale
dimension $d_{\cal U}$ and the unparticle coupling
$\lambda^{ff'}_{\rm K}$(K=V, A, S, P). Present experimental upper
bounds on ${\rm Br}(\mu\rightarrow e\gamma)$, ${\rm CR}(\mu-e,{\rm
Ti})$ and ${\rm CR}(\mu-e,{\rm Au})$ put stringent constraints on
the parameters of unaprticle physics. The scale dimensions $d_{\cal
U}$ around 2 are favored for the unparticle scale $\Lambda_{\cal U}$
of ${\cal O}(10\,{\rm TeV})$ and the unparticle coupling of ${\cal
O}(10^{-3})$. ${\rm CR}(\mu-e,{\rm Nuclei})$ is proportional to
$\rm{Z^4_{eff}A^2/Z}$ for the pure vector and scalar couplings
between unparticle and SM fermions, this peculiar atomatic number
dependence can be used to distinguish unparticle from other
theoretical models.

\vskip 0.5cm

%PACS numbers: 12.39.-x,13.20.Gd, 13.25.Gv,14.40.Lb

\end{abstract}
\maketitle
\section{introduction}
Scale invariance proves to be a very powerful concept in physics. At
low energy, the scale invariance is explicitly broken by the masses
of particles, and it is manifestly broken by the Higgs potential in
the standard model(SM). However, there may exist a scale invariant
sector at a much higher scale, e.g., above TeV scale. Motivated by
the Banks-Zaks theory\cite{Banks:1981nn}, recently Georgi suggested
that a scale invariant sector with nontrivial infrared fixed-point
may appear, which weakly couples to the
SM\cite{Georgi:2007ek,Georgi:2007si}. At low energy scale, this
sector is matched onto the so called "unparticle" with non-integral
scale dimension $d_{\cal U}$. For simplicity, most literatures so
far have assumeed that scale invariance remains until the low energy
scale.

Unparticle is very peculiar from the view of particle physics, it
looks like a non-integral number $d_{\cal U}$ of invisible massless
particles, this leads to peculiar energy and momentum distributions,
through which unparticle may be detected in high energy colliders
\cite{Georgi:2007ek}. Unparticle doesn't have a definite invariant
mass, instead a continuous mass spectrum, which can be represented
by an infinite tower of massive particles from the perspective of
particle physics\cite{Stephanov:2007ry}. Moreover, the unparticle
two-point correlation function has an unusual phase in the time-like
region, which can produce interesting interference patterns between
the the amplitude of S-channel unparticle exchange and that of SM
processes\cite{Georgi:2007si}.

Despite of the complexities of the scale invariant sector, we can
use the effective theory to deal with its low energy behavior. The
unparticle operator can have different lorentz structures: scalar
${\cal O}_{\cal U}$, vector ${\cal O}^{\,\mu}_{\cal U}$, tensor
${\cal O}^{\,\mu\nu}_{\cal U}$ or spinor. However, so far there is
no principle to constrain the interactions between the SM fields and
unparticle. The rich unparticle phenomenological implications have
been extensively
studied\cite{Georgi:2007ek,Georgi:2007si,Stephanov:2007ry,CKY,LZ,CH12,chengeng,Ding_Yan,Y_Liao,Aliev,Catterall,Li,Lu,Fox,Greiner,Davou,CGM,XG,MR,Zhou,FFRS,Bander,Rizzo,GN,Zwicky,Kikuchi,M_Giri,Huang,Krasnikov,Lenz,C_Ghosh,Zhang,Nakayama,Deshpande,DEQ,Neubert,Hannestad,
Das,Bhattacharyya,Majumdar,Alan-Pak,FD,Gogoladze:2007jn,Hur:2007cr,Anchordoqui:2007dp,Majhi:2007tu,McDonald:2007bt,Kumar:2007af,Das:2007cc}
in particle physics, astrophysics, cosmology, gravity and so on.

In the minimal version of SM, lepton flavor violating(LFV)
interactions are strictly forbidden. In the minimal extension of SM
in order to accomodate the present data on neutrino masses and
mixings, the LFV processes, such as $\ell_i\rightarrow\ell_j\gamma
(i\neq j)$ and $\mu^{-}\rightarrow e^{-}e^{+}e^{-}$ are very
strongly suppressed due to tiny neutrino masses and unitarity of the
mixing matrix(MNS matrix). In particular, the branching ratio for
$\mu\rightarrow e\gamma$ amounts to at most $10^{-54}$, to be
compared with the present experimental upper bound
$1.2\times10^{-11}$\cite{pdg}. However, most extensions of the SM
predict LFV, and some of them predict LFV at much higher rates,
which may be in conflict with the existing experimental bounds. LFV
provides an unique insight into the nature of new physics beyond
SM\cite{Kuno:1999jp}, and various LFV processes have been considered
in many scenarios of new physics beyond SM, such as the see-saw
model with or without GUT\cite{Cheng:1976uq},
supersymmetry\cite{Ellis}, $\rm{Z'}$ model\cite{Nardi:1992nq} and so
on. Three kinds of LFV processes are usually discussed: LFV
radiative decays $\ell_i\rightarrow\ell_j\gamma\,(i\neq j)$,
$\mu\rightarrow 3e$ like processes and $\mu-e$ conversion in nuclei.
Unparticle induced $\mu^{-}\rightarrow e^{-}e^{+}e^{-}$ and other
cross symmetry related processes such as $e^{+}+e^{-}\rightarrow
e^{+}+\mu^{-}$ have been considered\cite{Aliev,Lu,CGM}. In this work
, we will consider LFV radiative decay $\mu\rightarrow e\gamma$ and
$\mu-e$ conversion in nuclei.

Besides the great theoretical interests, there has been a lot of
theoretical efforts in detecting LFV processes at CERN LEP and
B-factories. The current experimental bound on the LFV radiative
decay $\mu\rightarrow e\gamma$ is as follows\cite{pdg}
\begin{eqnarray}
\label{1}&&{\rm Br}(\mu\rightarrow
e\gamma)<1.2\times10^{-11}\,,~~~{\rm CL}=90\%
\end{eqnarray}
For $\mu-e$ conversion in heavy nuclei, the most stringent
constraints arise for Titanium and Gold, respectively with
$\rm{CR(\mu-e,Ti)<4.3\times10^{-12}}$\cite{Dohmen:1993mp} and
$\rm{CR(\mu-e,Au)<7\times10^{-13}}$\cite{Bertl:2001fu}. Several
experiments have been designed to explore LFV with much higher
sensitivity than presently available. In particular, the MEG
experiment at PSI will detect $\mu\rightarrow e\gamma$ down to the
$10^{-13}-10^{-14}$ level in the very near future\cite{Ritt:2006cg}.
Concerning the challenging $\mu-e$ conversion in heavy nuclei, the
J-PARC experiment PRISM/PRIME is expected to reach a sentsitivity of
${\cal O}(10^{-18})$\cite{mueconv}, i.e. an improvement by six
orders of magnitude relative to the present upper bound.

Motivated by the future considerable progress in experimental
measuements, studying $\mu\rightarrow e\gamma$ and $\mu-e$
conversion in unparticle physics are of great theoretical interests.
The paper is organized as follows. In Section II, we review the
baisc aspects of unparticle physics. In Section III, we calculate
the LFV radiative decays $\mu\rightarrow e\gamma$. $\mu-e$
conversion rates in nuclei are considered in Section IV. Finally we
present our conclusions and some discussions.

\section{the model}

As was suggested by Georgi\cite{Georgi:2007ek}, we shall assume that
at a very high energy scale, the world consists of the SM sector and
the so-called Banks-Zaks (${\cal BZ}$) sector with non-trivial
infrared(IR) fixed point, and the two sectors interact with each
other via the exchange of particles with a large mass scale ${\rm
M}_{\cal U}>>1\rm{TeV}$. Below the scale ${\rm M}_{\cal U}$, the
interactions between these two sectors may be described by the
effective non-renormalizable Lagrangian,
\begin{equation}
\label{2} \frac{1}{M^{d_{\cal BZ}+d_{SM}-4}_{\cal U}}\;{\cal
O}_{SM}{\cal O}_{\cal BZ}
\end{equation}
which is analogous to the four-fermion interactions in SM, where
${\cal O}_{SM}$ and ${\cal O}_{\cal BZ}$ are respectively local
operators constructed from the SM fields and the ${\cal BZ}$ fields.
The ${\cal BZ}$ theory has IR fixed point around an energy scale
$\Lambda_{\cal U}\sim 1\rm{TeV}$, below this scale, the ${\cal BZ}$
sector undergoes dimensional transmutation and the scale invariant
unparticle sector emerges. The ${\cal BZ}$ operator ${\cal O}_{\cal
BZ}$ matches onto the unparticle operator ${\cal O}_{\cal U}$, and
the interactions between the unparticle and the SM fields generally
have the form
\begin{equation}
\label{3}{\cal L}_{eff}=\frac{\lambda}{\Lambda_{\cal U}^{d_{\cal
U}+d_{SM}-4}}\;{\cal O}_{SM}{\cal O}_{\cal U}
\end{equation}
where $\lambda=C_{\cal U}\,(\frac{\Lambda_{\cal U}}{M_{\cal
U}})^{d_{\cal BZ}+d_{SM}-4}$ and $C_{\cal U}$ is the Wilson-like
coefficient function. The lowest order effective interactions
between the unparticle and the SM fermion fields are as follows
\begin{equation}
\label{4}{\cal L}_{int}=\frac{\lambda^{ff'}_{\rm
V}}{\Lambda^{d_{\cal U}-1}_{\cal U}}\overline{f}\gamma_{\mu}f'{\cal
O}^{\,\mu}_{\cal U}+\frac{\lambda^{ff'}_{\rm A}}{\Lambda^{d_{\cal
U}-1}_{\cal U}}\overline{f}\gamma_{\mu}\gamma_5f'{\cal
O}^{\,\mu}_{\cal U}+\frac{\lambda^{ff'}_{\rm S}}{\Lambda^{d_{\cal
U}-1}_{\cal U}}\overline{f}f'{\cal O}_{\cal
U}+\frac{\lambda^{ff'}_{\rm P}}{\Lambda^{d_{\cal U}-1}_{\cal
U}}\overline{f}i\gamma_5f'{\cal O}_{\cal U}
\end{equation}
Here $f$ and $f'$ denote SM fermions(leptons or quarks), and they
should have same electric charges. We note that both the third and
the fourth term are absent, if we require that the effective
Lagragian ${\cal L}_{int}$ is consistent with the SM symmetry with
unparticle being SM singlet. The unparticle operators have been set
to be hermitian, and ${\cal O}^{\,\mu}_{\cal U}$ is assumed to be
transverse $\partial_{\mu}{\cal O}^{\,\mu}_{\cal U}=0$. The
couplings between the SM fermion fields and unparticle are quite
arbitrary, it can be flavor conserving or changing. Moreover, there
is no any correlation in the transitions among three generations for
flavor changing processes. In Ref.\cite{chengeng}, the authors
introduced ${\cal BZ}$ charges for the SM particles at very high
energy scale, then tree level flavor changing neutral current(FCNC)
can be induced by rediagonalizations of the SM fermion mass
matrices. Under the Fritzsch ansatz of the mass matrices, the FCNC
effects were found to associated with the mass ratios
$\sqrt{{m_im_j}/{m^2_3}}$. Scale invariance fixes the two-point
correlation function of unparticle, by dispersion relation, the
two-point correlation function is determined to
be\cite{Georgi:2007si,CKY}
\begin{equation}
\label{5}\int d^4x\,e^{iP\cdot x}\langle 0|T({\cal O}_{\cal
U}(x){\cal O}^{\,\dagger}_{\cal U}(0))|\,0\rangle=\frac{iA_{d_{\cal
U}}}{2\pi}\int_0^{\infty}\frac{s^{\,d_{\cal
U}-2}}{P^2-s+i\epsilon}ds=\frac{iA_{d_{\cal U}}}{2\sin(d_{\cal
U}\,\pi)}(-P^2-i\epsilon)^{d_{\cal U}-2}
\end{equation}
where the normalization factor $A_{d_{\cal U}}$ is chosen to be
\begin{equation}
\label{6}A_{d_{\cal U}}=\frac{16\pi^{5/2}}{(2\pi)^{2\,d_{\cal
U}}}\frac{\Gamma(d_{\cal U}+\frac{1}{2})}{\Gamma(d_{\cal
U}-1)\Gamma(2d_{\cal U})}
\end{equation}
and the complex function $(-P^2-i\epsilon)^{d_{\cal U}-2}$ is
defined to be
\begin{equation}
\label{7}(-P^2-i\epsilon)^{d_{\cal U}-2}=\left\{\begin{array}{cr}
(|P^2|-i\epsilon)^{d_{\cal U}-2}\,,&P^2<0\\
(|P^2|+i\epsilon)^{d_{\cal U}-2}e^{-id_{\cal
U}\,\pi}\,,&P^2>0\end{array}\right.
\end{equation}
In Eq.(\ref{5}) the unparticle operator is scalar, it is
straightforwardly to generalize to the vector unparticle operator
${\cal O}^{\,\mu}_{\cal U}$
\begin{equation}
\label{8}\int d^4x\,e^{iP\cdot x}\langle 0|T({\cal O}^{\,\mu}_{\cal
U}(x){\cal O}^{\,\nu\dagger}_{\cal
U}(0))|\,0\rangle=\frac{iA_{d_{\cal U}}}{2\sin(d_{\cal
U}\,\pi)}(-P^2-i\epsilon)^{d_{\cal
U}-2}\,(-g^{\mu\nu}+P^{\mu}P^{\nu}/P^2)
\end{equation}
We note that the dispersion representation of the unparticle
two-point correlation function is very useful, if unparticle appear
in the loop, e.g. the unparticle induced lepton anomalous magnetic
momentum and LFV radiative decay $\mu\rightarrow e\gamma$ in the
following.

\section{LFV radiative decays} %$\ell_i\rightarrow\ell_j\gamma$

\vskip0.5cm
\begin{figure}[hptb]
\subfigure[]{\label{subfig:a}
\begin{minipage}[t]{0.33\textwidth}
\includegraphics*[87pt,720pt][203pt,800pt]{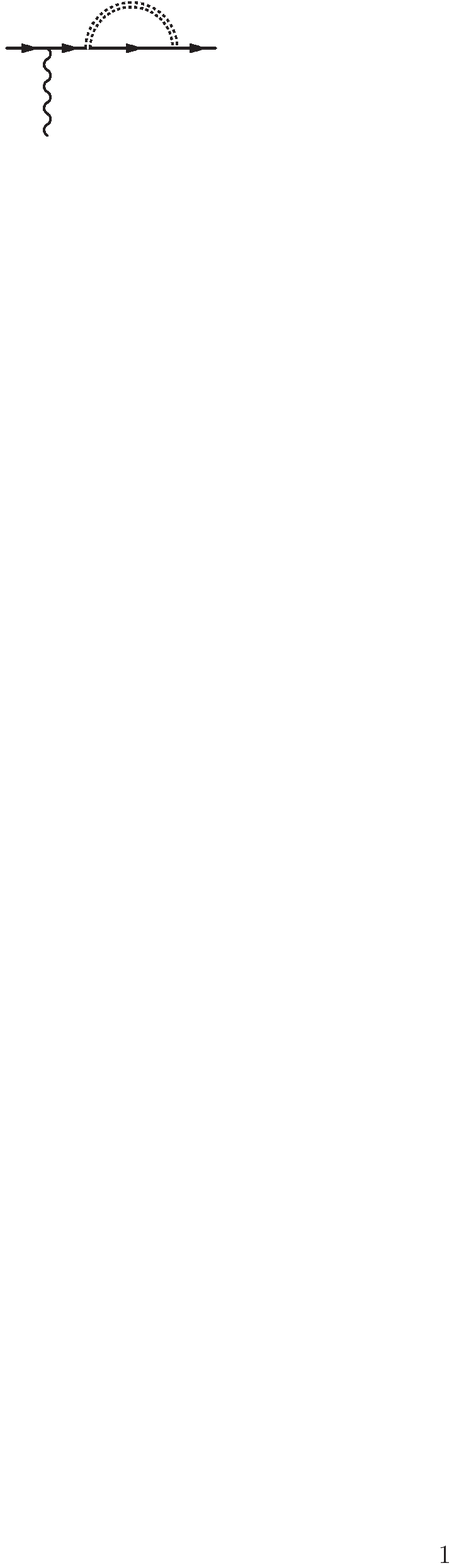}
\end{minipage}}%
\subfigure[]{\label{subfig:b}
\begin{minipage}[t]{0.33\textwidth}
\includegraphics*[87pt,720pt][203pt,800pt]{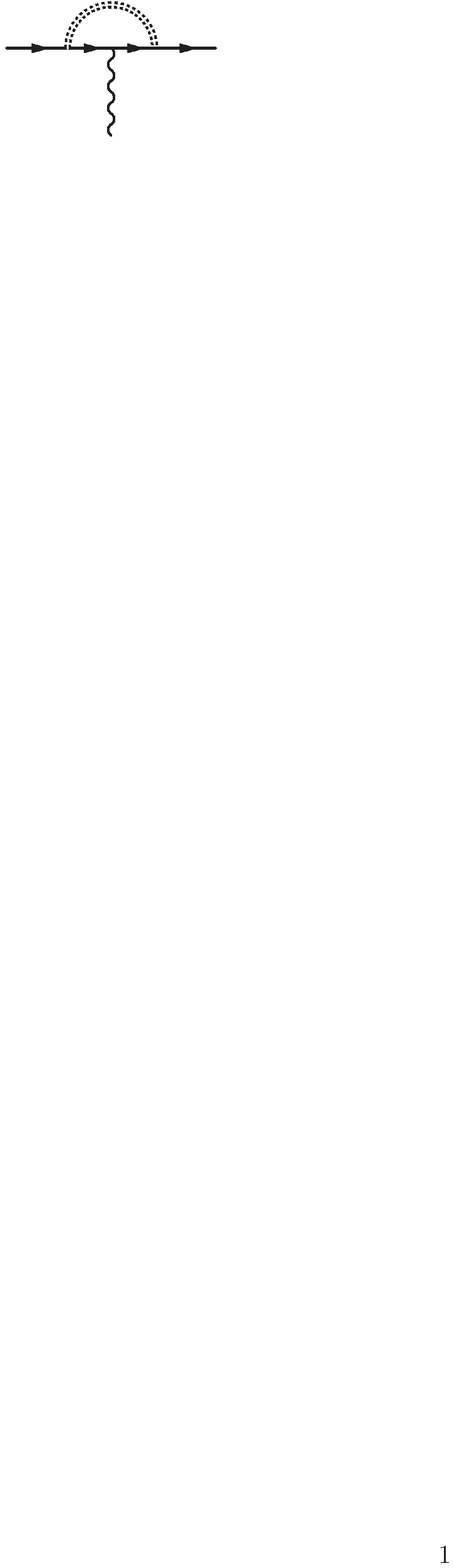}
\end{minipage}}%
\subfigure[]{\label{subfig:c}
\begin{minipage}[t]{0.33\textwidth}
\includegraphics*[87pt,720pt][203pt,800pt]{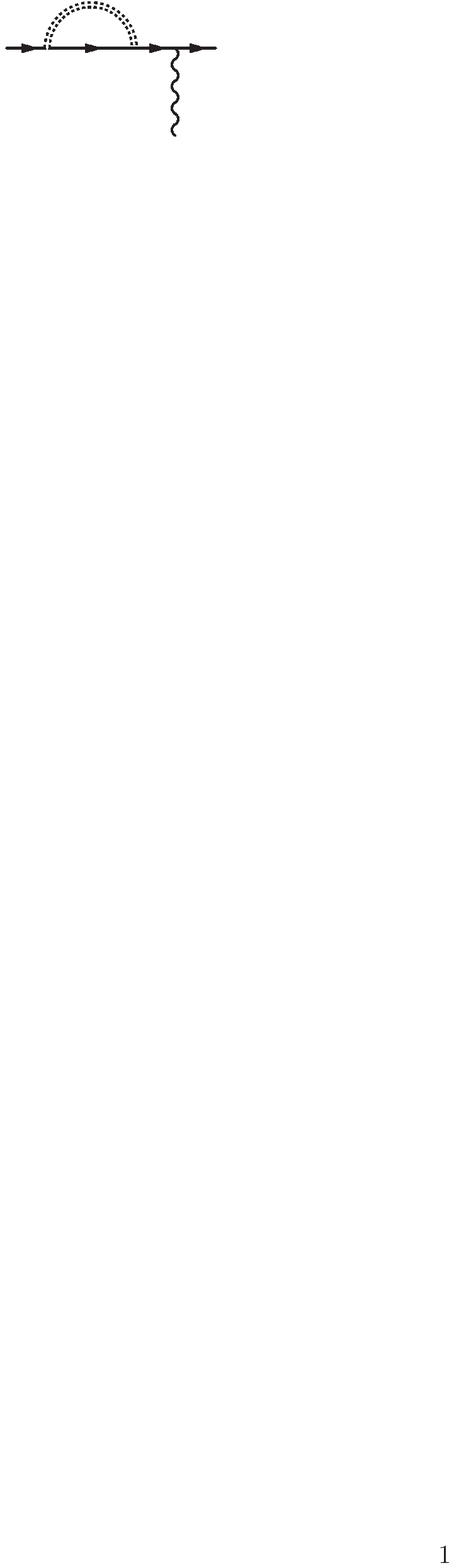}
\end{minipage}} \caption{The Feynman diagrams contributing to $\mu\rightarrow e\gamma$} \label{LFV}
\end{figure}
The diagrams for the LFV $\mu\rightarrow e\gamma$ are shown in
Fig.\ref{LFV}. Generally, the amplitude for $\mu\rightarrow e\gamma$
can be written as
\begin{equation}
\label{9}{\cal M}(\mu\rightarrow
e\gamma)=\varepsilon^{\mu*}\overline{u}_e(p_e)[iq^{\nu}\sigma_{\mu\nu}({\rm
A}+{\rm B}\gamma_5)+\gamma_{\mu}({\rm C}+{\rm
D}\gamma_5)+q_{\mu}({\rm E}+{\rm F}\gamma_5)]u_{\mu}(p_{\mu})
\end{equation}
where $q_{\mu}$ and $\varepsilon^{\mu}$ are respectively the photon
momentum and polarization, ${\rm A,B,...,F}$ are invariant
amplitudes. The electromagnetic gauge invariance requires the above
amplitude is invariant under
$\varepsilon^{\mu}\rightarrow\varepsilon^{\mu}+q^{\mu}$, then we
have,
\begin{equation}
\label{10}{\rm C=D=0}
\end{equation}
Since the photon is on shell $q^2=0$ and transverse
$\varepsilon^{\mu}q_{\mu}=0$, $\mu\rightarrow e\gamma$ is a magnetic
transition
\begin{equation}
\label{11}{\cal M}(\mu\rightarrow
e\gamma)=\varepsilon^{\mu*}\overline{u}_e(p_e)[iq^{\nu}\sigma_{\mu\nu}({\rm
A}+{\rm B}\gamma_5)]u_{\mu}(p_{\mu})
\end{equation}
It is easy to calculate of the corresponding radiative decay width
\begin{equation}
\label{12}\Gamma(\mu\rightarrow
e\gamma)={\rm\frac{m^3_{\mu}}{8\pi}(|A|^2+|B|^2)}
\end{equation}
where we have neglected the final state electron mass. Using
$\Gamma(\mu\rightarrow e\overline{\nu}_e\nu_{\mu})={\rm
m^5_{\mu}G^2_F/192\pi^3}$, here ${\rm G_F}$ is the Fermi constant,
this can be converted into the branching ratio
\begin{equation}
\label{add1}{\rm Br}(\mu\rightarrow
e\gamma)=\frac{\Gamma(\mu\rightarrow e\gamma)}{\Gamma(\mu\rightarrow
e\overline{\nu}_e\nu_{\mu})}=\frac{24\pi^2}{\rm
m^2_{\mu}G^2_F}\,(|A|^2+|B|^2)
\end{equation}
We note that the couplings between unparticle and photon such as
${\cal O}^{\,\mu\alpha}_{\cal U}{\cal O}_{{\cal
U}\,\nu\alpha}F_{\mu}^{\,\nu}$ also contribute to $\mu\rightarrow
e\gamma$, However, its contribution is highly suppressed compared
with those in Fig.\ref{LFV}. Using the dispersion representation of
the unparticle two-point correlation function, it is
straightforward, albeit some lengthy, to work out these unparticle
induced radiative decay $\mu\rightarrow e\gamma$ amplitude. In fact,
we only need to consider Fig.\ref{subfig:b}, since the contribution
of Fig.\ref{subfig:a} and Fig.\ref{subfig:c} are proportional to
$\varepsilon^{\mu*}\overline{u}_e(p_e)\gamma_{\mu}u_{\mu}(p_{\mu})$
or
$\varepsilon^{\mu*}\overline{u}_e(p_e)\gamma_{\mu}\gamma_5u_{\mu}(p_{\mu})$.

\begin{eqnarray}
\nonumber A_{S}&=&\frac{eA_{d_{\cal U}}}{4\sin(d_{\cal
U}\pi)}\frac{-i}{(4\pi)^2}\sum_{a=e,\,\mu,\,\tau}\frac{\lambda^{ea}_S\lambda^{a\mu}_S}{(\Lambda^2_{\cal
U})^{d_{\cal
U}-1}}\int_0^1dxdydz\,\delta(x+y+z-1)[2xzm_e+2yzm_{\mu}\\
\label{13}&&+2(x+y)m_a]\,e^{-i(d_{\cal
U}-2)\pi}[xzm^2_{e}+yzm^2_{\mu}-(x+y)m^2_a]^{d_{\cal
U}-2}z^{1-d_{\cal U}}
\end{eqnarray}
\begin{eqnarray}
\nonumber A_{P}&=&\frac{eA_{d_{\cal U}}}{4\sin(d_{\cal
U}\pi)}\frac{-i}{(4\pi)^2}\sum_{a=e,\,\mu,\,\tau}\frac{\lambda^{ea}_P\lambda^{a\mu}_P}{(\Lambda^2_{\cal
U})^{d_{\cal
U}-1}}\int_0^1dxdydz\,\delta(x+y+z-1)[2xzm_{e}+2yzm_{\mu}\\
\label{14}&&-2(x+y)m_a]\,e^{-i(d_{\cal
U}-2)\pi}[xzm^2_{e}+yzm^2_{\mu}-(x+y)m^2_a]^{d_{\cal
U}-2}z^{1-d_{\cal U}}
\end{eqnarray}
\begin{eqnarray}
\nonumber A_{V}&=&\frac{eA_{d_{\cal U}}}{4\sin(d_{\cal
U}\pi)}\frac{-i}{(4\pi)^2}\sum_{a=e,\,\mu,\,\tau}\frac{\lambda^{ea}_V\lambda^{a\mu}_V}{(\Lambda^2_{\cal
U})^{d_{\cal
U}-1}}\int_0^1dxdydz\,\delta(x+y+z-1)\{\,[-4z(1-x)m_{e}\\
\nonumber&&-4z(1-y)m_{\mu}+8zm_a]\,e^{-i(d_{\cal
U}-2)\pi}[xzm^2_{e}+yzm^2_{\mu}-(x+y)m^2_a]^{d_{\cal
U}-2}z^{1-d_{\cal
U}}+\\
\nonumber&&2\,[2y(m_a-m_{e})+2x(m_a-m_{\mu})-(1+z)m_a+2z(xm_{e}+ym_{\mu})+z(1-x)m_{e}\\
\nonumber&&+x(1-z)m_{e}+z(1-y)m_{\mu}+y(1-z)m_{\mu}]\,e^{-i(d_{\cal
U}-2)\pi}[xzm^2_{e}+yzm^2_{\mu}-(x+y)m^2_a]^{d_{\cal
U}-2}\\
\nonumber&&\times z^{2-d_{\cal U}}/(2-d_{\cal
U})+[2y(m_a-m_{e})(xzm^2_{e}+(1-y)(1-z)m^2_{\mu}+ym_am_{\mu})+\\
\nonumber&&2x(m_a-m_{\mu})(yzm^2_{\mu}+(1-x)(1-z)m^2_{e}+xm_{e}m_a)+2xy(m_{e}+m_{\mu})(m_a-m_{e})\\
\label{15}&&\times(m_a-m_{\mu})]\,e^{-i(d_{\cal
U}-3)\pi}[xzm^2_{e}+yzm^2_{\mu}-(x+y)m^2_a]^{d_{\cal
U}-3}z^{2-d_{\cal U}}\}
\end{eqnarray}
\begin{eqnarray}
\nonumber A_{A}&=&\frac{eA_{d_{\cal U}}}{4\sin(d_{\cal
U}\pi)}\frac{-i}{(4\pi)^2}\sum_{a=e,\,\mu,\,\tau}\frac{\lambda^{ea}_A\lambda^{a\mu}_A}{(\Lambda^2_{\cal
U})^{d_{\cal
U}-1}}\int_0^1dxdydz\,\delta(x+y+z-1)\{\,[-4z(1-x)m_{e}\\
\nonumber&&-4z(1-y)m_{\mu}-8zm_a]\,e^{-i(d_{\cal
U}-2)\pi}[xzm^2_{e}+yzm^2_{\mu}-(x+y)m^2_a]^{d_{\cal
U}-2}z^{1-d_{\cal
U}}+\\
\nonumber&&2\,[-2y(m_a+m_{e})-2x(m_a+m_{\mu})+(1+z)m_a+2z(xm_{e}+ym_{\mu})+z(1-x)m_{e}\\
\nonumber&&+x(1-z)m_{e}+z(1-y)m_{\mu}+y(1-z)m_{\mu}]\,e^{-i(d_{\cal
U}-2)\pi}[xzm^2_{e}+yzm^2_{\mu}-(x+y)m^2_a]^{d_{\cal
U}-2}\\
\nonumber&&\times z^{2-d_{\cal U}}/(2-d_{\cal
U})+[-2y(m_a+m_{e})(xzm^2_{e}+(1-y)(1-z)m^2_{\mu}-ym_am_{\mu})\\
\nonumber&&-2x(m_a+m_{\mu})(yzm^2_{\mu}+(1-x)(1-z)m^2_{e}-xm_{e}m_a)+2xy(m_{e}+m_{\mu})(m_a+m_{e})\\
\label{16}&&(m_a+m_{\mu})]\,e^{-i(d_{\cal
U}-3)\pi}[xzm^2_{e}+yzm^2_{\mu}-(x+y)m^2_a]^{d_{\cal
U}-3}z^{2-d_{\cal U}}\}
\end{eqnarray}

\begin{figure}[hptb]
\centering
\includegraphics*[width=0.8\textwidth]{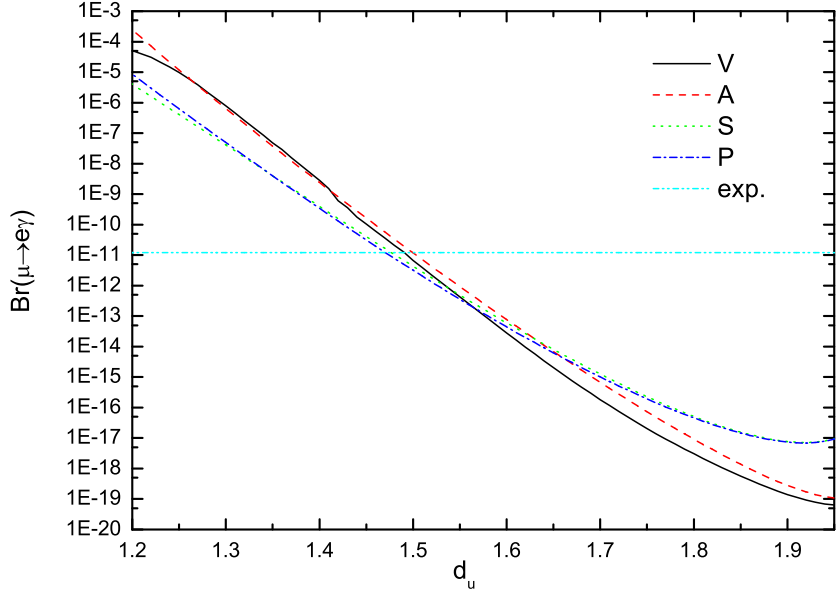}
\caption{\label{mu-egamma}Variation of the branching ratios ${\rm
Br}(\mu\rightarrow e\gamma)$ with the scale dimension $d_{\cal U}$.
V, A, S and P respectively denote the branching ratios for the pure
vector, axial vector, scalar and pseudoscalar couplings between
unparticle and SM fermions. The horozontal line indicates the
present experimental bounds for ${\rm Br}(\mu\rightarrow e\gamma)$.
We have taken $\lambda_{\rm V}=\lambda_{\rm A}=\lambda_{\rm
S}=\lambda_{\rm P}=0.001$, $\kappa=3$ and $\Lambda_{\cal U}=10{\,\rm
TeV}$. }
\end{figure}

where the subscript denotes the contribution from the corresponding
interactions between unparticle and the SM fermions, $B_S$, $B_{P}$,
$B_V$ and $B_{A}$ equal zero. If both vector coupling and axial
vector coupling between the unparticle and fermions(or scalar
coupling and pseudoscalar coupling) exist simultaneously, $B$ would
be non-zero. In Eq.(\ref{13})-Eq.(\ref{16}), there is the factor
$[xzm^2_{e}+yzm^2_{\mu}-(x+y)m^2_a]^{d_{\cal U}-2}$ with
$a=e,\mu,\tau$ ( or $[xzm^2_{e}+yzm^2_{\mu}-(x+y)m^2_a]^{d_{\cal
U}-3}$). It is well-defined if $xzm^2_{e}+yzm^2_{\mu}-(x+y)m^2_a>0$,
whereas $[xzm^2_{e}+yzm^2_{\mu}-(x+y)m^2_a]^{d_{\cal
U}-2}=\exp(i(d_{\cal
U}-2)\pi)[-xzm^2_{e}-yzm^2_{\mu}+(x+y)m^2_a]^{d_{\cal U}-2}$ if
$xzm^2_{e}+yzm^2_{\mu}-(x+y)m^2_a<0$. Note that $A_V$ and $A_A$ are
computed for a transverse vector unparticle operator ${\cal
O}^{\,\mu}_{\cal U}$, both $g^{\mu\nu}$ and $P^{\mu}P^{\nu}/P^2$
parts in the unparticle two-point correlation function contribute to
the decay amplitude.

In Fig.\ref{mu-egamma} we present the variation of the branching
ratio ${\rm Br}(\mu\rightarrow e\gamma)$ as a function of the scale
dimension $d_{\cal U}$ respectively for the pure vector, axial
vector, scalar, pseudoscalar couplings between unparticle and the SM
fermions. For simplicity, we assume that the unparticle couplings
with the SM fermions are universal
\begin{equation}
\label{add2}\lambda^{ff'}_{\rm K}=\left\{\begin{array}{cr}
\lambda_{\rm K}\,,&f\neq f'\\
\kappa\lambda_{\rm K}\,,&f=f'\end{array}\right.
\end{equation}
where $\kappa>1$ and K=V, A, S, P for vector, axial vector, scalar,
pseudoscalar couplings respectively .

\begin{figure}[hptb]
\begin{center}
\includegraphics*[width=0.8\textwidth]{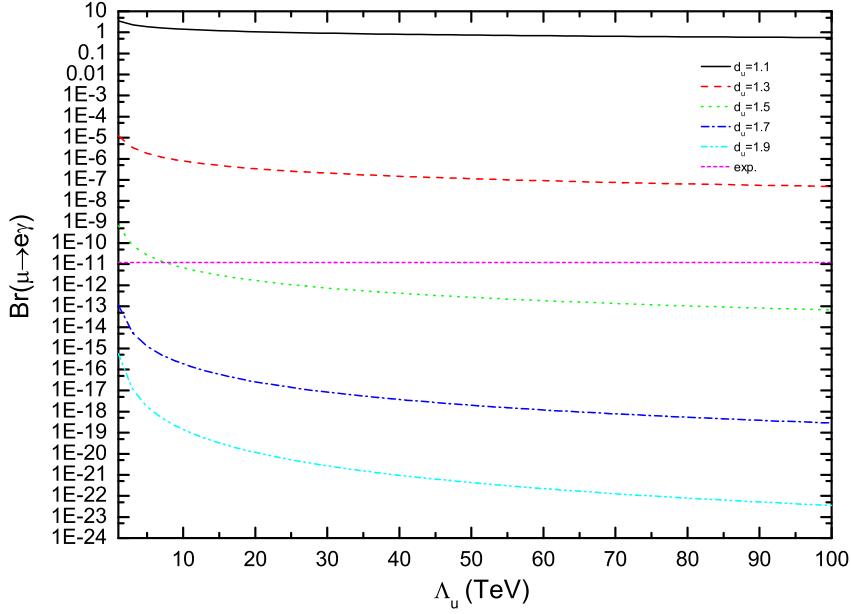}
\caption{\label{Brmue-Lambda}${\rm Br}(\mu\rightarrow e\gamma)$ as a
function of the unparticle scale $\Lambda_{\cal U}$ for various
scale dimension $d_{\cal U}$ in the pure vector coupling case. The
horozontal line indicates the present experimental bound for ${\rm
Br}(\mu\rightarrow e\gamma)$. We have taken $\lambda_{\rm V}=0.001$,
$\kappa=3$. }
\end{center}
\end{figure}

As we can see from Fig.\ref{mu-egamma}, the branching ratio ${\rm
Br}(\mu\rightarrow e\gamma)$ decreases with $d_{\cal U}$ in the
considered range, and it strongly depends on the scale dimension
$d_{\cal U}$. There is little difference between ${\rm
Br}(\mu\rightarrow e\gamma)$ in the pure vector coupling case and
that in the pure axial vector coupling case. The same is true for
the pure scalar coupling and the pseudoscalar coupling. From
Eq.(\ref{add1}) and Eq.(\ref{13})-Eq.(\ref{16}), we can see ${\rm
Br}(\mu\rightarrow e\gamma)$ is proportional to $1/(\Lambda^2_{\cal
U})^{2d_{\cal U}-2}$. The $\Lambda_{\cal U}$ dependence of ${\rm
Br}(\mu\rightarrow e\gamma)$ for the pure vector coupling case is
shown in Fig.\ref{Brmue-Lambda}.

From Fig.\ref{Brmue-Lambda}, we find that ${\rm Br}(\mu\rightarrow
e\gamma)$, for $d_{\cal U}=1.1$ or 1.3 and other input parameters in
that figure, is clearly above its present experimental upper bound.
The important conclusion from Fig.\ref{mu-egamma} and
Fig.\ref{Brmue-Lambda} is that the present experimental data on
${\rm Br}(\mu\rightarrow e\gamma)$ favors the scale dimension
$d_{\cal U}$ close to 2 for $\Lambda_{\cal U}$ of ${\cal O}$(10 TeV)
and the unparticle couplings of ${\cal O}(10^{-3})$.

\section{$\mu-e$ conversion in nuclei}

\vskip0.5cm
\begin{figure}[hptb]
\begin{center}
\includegraphics*[200pt,699pt][406pt,785pt]{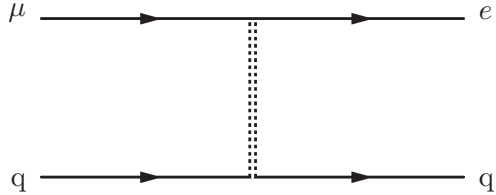}
\caption{\label{mueconv}Feymann diagram for $\mu-e$ conversion in
nuclei}
\end{center}
\end{figure}
The $\mu-e$ conversion in nuclei is described by the Feynman diagram
presented in Fig.\ref{mueconv}, it means the following exotic
process
\begin{equation}
\label{17}{\rm \mu^{-}+(A,Z)}\rightarrow e^{-}+{\rm(A,Z)}
\end{equation}
It violates the conservation of lepton flavor number $L_{e}$ and
$L_{\mu}$ by one unit, but conserve the total lepton number $L$. The
$\mu-e$ conversion rate is usually expressed by
\begin{equation}
\label{18}{\rm{CR}}(\mu-e,X)=\frac{\Gamma(\mu +X\rightarrow e+X
)}{\Gamma(\mu+X\rightarrow{\rm capture})}
\end{equation}
where $\Gamma(\mu+X\rightarrow \rm{capture})$ is the $\mu$ capture
rate of the nuclei $X$. A very detailed calculation of the $\mu-e$
conversion rate in various nuclei has been performed
in\cite{Kitano:2002mt}, using the methods developed by Czarnecki et
al. \cite{Czarnecki:1998iz}. It has been emphasized in
\cite{Kitano:2002mt} that the atomic number dependence of the
conversion rate can be used to distinguish between different
theoretical models of LFV.

We will calculated the $\mu-e$ conversion rates in nuclei using the
general model-independent formulae of both \cite{Kitano:2002mt} and
\cite{Czarnecki:1998iz}. For the nucleon numbers relevant for
$\mu-e$ conversion experiments, the rate for the coherent process
dominate over the incoherent excitations of the nuclear system, and
the rate of the coherent conversion process over the nocoherent ones
is enhanced by a factor approximately equal to the number of
nucleons in nucleus. Explicit calculations based on nuclear models
\cite{Kosmas:1990tc} show that the ratio between the coherent rate
and the total $\mu-e$ conversion rate for nuclei as $^{48}$Ti can be
as large as 90\%.

For coherent $\mu-e$ conversion, only vector coupling and scalar
coupling between the quarks and unparticle do contribute, and the
contributions of axial vector and pseudoscalar couplings are
negligible. For the pure vector coupling between SM fermions and
unparticle, the four fermion effective interaction, which describes
coherent $\mu-e$ conversion, is given by
\begin{equation}
\label{19}{\cal L}^{\rm V}_{\mu-e~\rm{conv}}=\lambda^{e\mu}_{\rm
V}\lambda^{qq}_{\rm V}\frac{A_{d_{\cal U}}}{2\sin(d_{\cal
U}\pi)}\frac{1}{\Lambda^2_{\cal U}}(\frac{-{\rm
q}^2}{\Lambda^2_{\cal U}})^{d_{\cal
U}-2}\,\overline{e}\gamma^{\mu}\mu\,\overline{q}\gamma_{\mu}q
\end{equation}
For the pure scalar coupling case,
\begin{equation}
\label{20}{\cal L}^{\rm S}_{\mu-e~\rm{conv}}=\lambda^{e\mu}_{\rm
S}\lambda^{qq}_{\rm S}\frac{A_{d_{\cal U}}}{2\sin(d_{\cal
U}\pi)}\frac{1}{\Lambda^2_{\cal U}}(\frac{-{\rm
q}^2}{\Lambda^2_{\cal U}})^{d_{\cal
U}-2}\,\overline{e}\mu\,\overline{q}q
\end{equation}
In Eq.(\ref{19}) and Eq.(\ref{20}), ${\rm q}^2$ is the momentum
transfer in the $\mu-e$ conversion process(${\rm
q}^2\simeq-m^2_{\mu}$), which is much smaller than the scale
associated with the structure of the nucleon, and we can neglect the
${\rm q}^2$ dependence in the nucleon form factors. The above
effective Lagrangian at the quark level is then converted to the
effective Lagrangian at the nucleon level, by means of the
approximate nucleon form factors\cite{Kuno:1999jp,Kitano:2002mt}.
The matrix elements of the quark current for the nucleon $N = p, n$
can be written as,
\begin{eqnarray}
\nonumber\langle
p|\overline{q}\,\Gamma_{\rm K}q|p\rangle&=&{\rm G}^{(q,\,p)}_{\rm K}\overline{p}\,\Gamma_{\rm K}p\\
\label{21}\langle n|\overline{q}\,\Gamma_{\rm K}q|n\rangle&=&{\rm
G}^{(q,\,n)}_{\rm K}\overline{n}\,\Gamma_{\rm K}n
\end{eqnarray}
where $\Gamma_{\rm K}=1,\gamma_{\mu}$ respectively for ${\rm
K=S,V}$. The numerical values of the relevant ${\rm G_{K}}$ are as
follows\cite{Kuno:1999jp}
\begin{eqnarray}
\nonumber&&{\rm G}^{(u,\,p)}_{\rm V}={\rm G}^{(d,\,n)}_{\rm V}=2,~~~{\rm G}^{(d,\,p)}_{\rm V}={\rm G}^{(u,\,n)}_{\rm V}=1,~~~{\rm G}^{(s,\,p)}_{\rm V}={\rm G}^{(s,\,n)}_{\rm V}=0\\
\label{22} &&{\rm G}^{(u,\,p)}_{\rm S}={\rm G}^{(d,\,n)}_{\rm
S}=5.1,~{\rm G}^{(d,\,p)}_{\rm S}={\rm G}^{(u,\,n)}_{\rm
S}=4.3,~{\rm G}^{(s,\,p)}_{\rm S}={\rm G}^{(s,\,n)}_{\rm S}=2.5
\end{eqnarray}
Under the approximation of equal proton and neutron densities in the
nucleus, and of non-relativistic muon wavefunction for the $1s$
state, the final formula for the $\mu-e$ conversion rate for the
pure vector coupling between SM fermions and unparticle, relative to
the muon capture rate, is given by
\begin{eqnarray}
\nonumber{\rm CR}(\mu-e,{\rm Nucleus})&=&\frac{{\rm p}_e{\rm
E}_e{\rm m}^3_{\mu}\alpha^3{\rm Z^4_{eff}}{\rm F^2_p}}{2\pi^2{\rm
Z}}\,[\lambda^{e\mu}_{\rm V}\lambda^{qq}_{\rm V}\frac{A_{d_{\cal
U}}}{2\sin(d_{\cal U}\pi)}\frac{1}{\Lambda^2_{\cal U}}(\frac{{\rm
m}_{\mu}^2}{\Lambda^2_{\cal U}})^{d_{\cal
U}-2}]^2|{\rm Z}\sum_q{\rm G}^{(q,\,p)}_{\rm V}\\
\label{23}&&+{\rm N}\sum_q{\rm G}^{(q,\,n)}_{\rm
V}|^{\,2}\frac{1}{\Gamma_{{\rm capt}}}
\end{eqnarray}
For pure scalar coupling case, it is
\begin{eqnarray}
\nonumber{\rm CR}(\mu-e,{\rm Nucleus})&=&\frac{{\rm p}_e{\rm
E}_e{\rm m}^3_{\mu}\alpha^3\rm{Z^4_{eff}}{\rm F^2_p}}{2\pi^2{\rm
Z}}\,[\lambda^{e\mu}_{\rm S}\lambda^{qq}_{\rm S}\frac{A_{d_{\cal
U}}}{2\sin(d_{\cal U}\pi)}\frac{1}{\Lambda^2_{\cal U}}(\frac{{\rm
m}_{\mu}^2}{\Lambda^2_{\cal U}})^{d_{\cal
U}-2}]^2|{\rm Z}\sum_q{\rm G}^{(q,\,p)}_{\rm S}\\
\label{24}&&+{\rm N}\sum_q{\rm G}^{(q,\,n)}_{\rm
S}|^{\,2}\frac{1}{\Gamma_{{\rm capt}}}
\end{eqnarray}

where ${\rm Z}$ and ${\rm N}$ are the numbers of proton and neutron
in nucleus, while ${\rm Z_{eff}}$ ia an effective atomic charge,
obtained by averaging the muon wavefunction over the nuclear
density\cite{Kosmas:1990tc}. ${\rm F_p}$ is the nuclear matrix
element and $\Gamma_{\rm capt}$ denotes the total muon capture rate.
${\rm m}_{\mu}$ is the muon mass, ${\rm p}_e$ and ${\rm E}_e$ is the
momentum and energy of the electron. Since $\sum_q{\rm
G}^{(q,\,p)}_{\rm V}=\sum_q{\rm G}^{(q,\,n)}_{\rm V}=3$ and
$\sum_q{\rm G}^{(q,\,p)}_{\rm S}=\sum_q{\rm G}^{(q,\,n)}_{\rm
S}=11.9$, the $\mu-e$ conversion rate is proportional to
$\rm{Z^4_{eff}A^2/Z}$ with the atomic number ${\rm A=Z+N}$, which
can distinguish unparticle from other theoretical models.

\begin{figure}[h]
\begin{center}
\includegraphics*[width=0.9\textwidth]{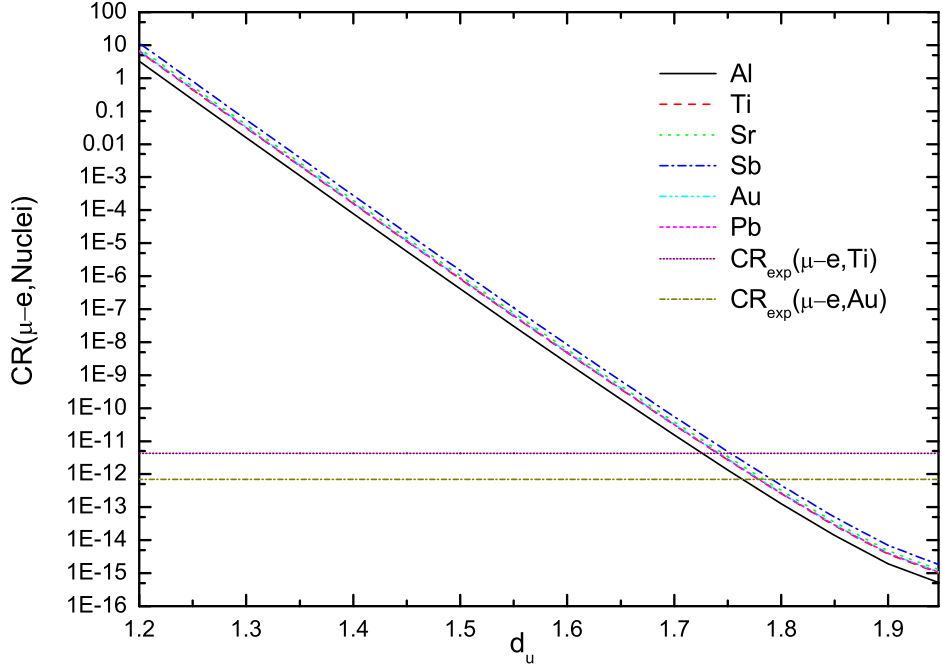}
\caption{\label{conv-du}$\mu-e$ conversion rates for various nuclei
as a function of the scale dimension $d_{\cal U}$ for the vector
coupling between unparticle and SM fermions. The horozontal lines
denote the present experimental bounds for ${\rm CR(\mu-e, Ti)}$ and
${\rm CR(\mu-e, Au)}$. We have taken $\lambda_V=0.001$, $\kappa=3$
and $\Lambda_{\cal U}=10{\rm TeV}$. }
\end{center}
\end{figure}

In Fig.\ref{conv-du}, we display the predicted $\mu-e$ conversion
rates for Al, Ti, Sr, Sb, Au and Pb as a function of the scale
dimension $d_{\cal U}$ in the case of vector coupling between
unparticle and SM fermions. The values of the relevant parameters
for these nuclei, ${\rm Z_{eff}}$, ${\rm F}_{\rm p}$ and
$\Gamma_{\rm capt}$ have been collected in Table
I\cite{Kitano:2002mt}. Here the universal couplings between
unparticle and SM fermions are assumed as we have done in
$\mu\rightarrow e\gamma$. We clearly see that the $\mu-e$ conversion
rates in nuclei ${\rm CR}(\mu-e,{\rm Nucleus})$ are sensitive to the
scale dimension $d_{\cal U}$, and they decrease with $d_{\cal U}$ as
well, which is obvious from Eq.(\ref{23}) and Eq.(\ref{24}), since
$({{\rm m}_{\mu}^2}/{\Lambda^2_{\cal U}})^{2d_{\cal U}-4}$ dominates
the $d_{\cal U}$ dependence in the plot range, and
$m^2_{\mu}/\Lambda^2_{\cal U}$ is a small quantity. Moreover, the
present experimental bound on ${\rm CR(\mu-e, Ti)}$ and ${\rm
CR(\mu-e, Au)}$ favor $d_{\cal U}$ near 2 for the input parameters
in this plot. The same conclusion has been found from LFV radiative
decay $\mu\rightarrow e\gamma$.

\begin{table}[hptb]
\caption{\label{mueconv parameter}The value of $\rm{Z_{eff}}$,
$\rm{F_p}$ and $\rm{\Gamma_{capt}}$ for various nucleis, which is
taken from \cite{Kitano:2002mt}.}
\begin{ruledtabular}
\begin{tabular}{cccc}
$\rm{_{Z}^{A}Nucleus}$ & $\rm{Z_{eff}}$ & $\rm{F_p}$ &
$\rm{\Gamma_{capt}(GeV)}$
\\\hline
$\rm{_{13}^{28}Al}$    & 11.5           & 0.64       &$4.64079\times10^{-19}$\\
$\rm{_{22}^{48}Ti}$    & 17.6           & 0.54       &$1.70422\times10^{-18}$\\
$\rm{_{38}^{80}Sr}$    & 25.0           & 0.39       &$4.61842\times10^{-18}$\\
$\rm{_{51}^{121}Sb}$   & 29.0           & 0.32       &$6.71711\times10^{-18}$\\
$\rm{_{79}^{197}Au}$   & 33.5           & 0.16       &$8.59868\times10^{-18}$\\
$\rm{_{82}^{207}Pb}$   & 34.0           & 0.15
&$8.84868\times10^{-18}$
\end{tabular}
\end{ruledtabular}
\end{table}

\section{summary}

Since LFV processes are sensitive probes to new physics beyond SM,
we have explored the peculiar aspects of unparticle physics in
$\mu\rightarrow e\gamma$ and $\mu-e$ conversion in nuclei, where
vector, axial vector, scalar, pseudoscalar couplings between
unparticle and SM fermions are considered. The difference between
the branching ratio ${\rm Br}(\mu\rightarrow e\gamma)$ in the pure
vector coupling case and that in the pure axial vector coupling case
is small, the same is true for scalar coupling and pseudoscalar
coupling. Only pure vector coupling and scalar coupling contribute
to $\mu-e$ conversion in nuclei, which is proportional to ${\rm
Z^4_{ eff}A^2/Z}$, which can be used to distinguish unparticle from
other theoretical models. Both ${\rm Br}(\mu\rightarrow e\gamma)$
and ${\rm CR}(\mu-e,{\rm Nuclei})$ are sensitive to the scale
dimension $d_{\cal U}$ and the unparticle coupling
$\lambda^{ff'}_{\rm K}$(K=V,A,S,P), and the present data on ${\rm
Br}(\mu\rightarrow e\gamma)$, ${\rm CR}(\mu-e,{\rm Ti})$ and ${\rm
CR}(\mu-e,{\rm Au})$ put stringent constraints on the parameters of
unparticle stuff. The scale dimensions $d_{\cal U}$ near 2 are
favored for the unparticle scale $\Lambda_{\cal U}$ of ${\cal
O}(10\,{\rm TeV})$ and the unparticle coupling of ${\cal
O}(10^{-3})$. The interactions between unparticle and SM fermions
can also lead to LFV $\mu\rightarrow e^{-}e^{+}e^{-}$ and cross
symmetry related processes such as $e^{+}+e^{-}\rightarrow
e^{+}+\mu^{-}$, detailed analyses of these processes have been
performed\cite{Aliev,Lu,CGM}. Future dedicated LFV measurments MEG
experiment and J-PARC experiment PRISM/PRIME would provide important
clues to understanding the nature of unparticle.

Unparticle associated with conformal hidden sector may exist, and it
has very distinctive phenomenologies. Unparticle may weakly couples
to the SM field so that we are able to explore the peculiar
properties of unparticle, However, whether observable effects can be
produced strongly depends on how weakly the unparticle interacts
with ordinary matter. So far there is no principle to constraint and
organize the interactions between the SM particles and unparticles,
therefore there are many freedoms in the present phenomenological
studies of unparticle. It would be enlightened and interesting to
build an explicit model, where hidden sector with strict or broken
scale invariance is realized and it connects to the SM fields via a
connector sector. These issues lie outside the scope of the this
work, and will be considered elsewhere\cite{ding}.

\section*{ACKNOWLEDGEMENTS}
\indent  We are grateful to Prof. Dao-Neng Gao for very helpful and
stimulating discussions, to Tzu-Chiang Yuan for correspondence. This
work is partially supported by National Natural Science Foundation
of China under Grant Numbers 90403021, and KJCX2-SW-N10 of the
Chinese Academy.


\begin{thebibliography}{99}
\bibitem{Banks:1981nn}
T.~Banks and A.~Zaks,
%``On The Phase Structure Of Vector-Like Gauge Theories With Massless
%Fermions,''
Nucl.\ Phys.\  B {\bf 196}, 189 (1982).
\bibitem{Georgi:2007ek}
H.~Georgi,
%``Unparticle Physics,''
Phys.\ Rev.\ Lett.\  {\bf 98}, 221601 (2007) [arXiv:hep-ph/0703260].
\bibitem{Georgi:2007si}
H.~Georgi,
%``Another Odd Thing About Unparticle Physics,''
Phys.\ Lett.\  B {\bf 650}, 275 (2007) [arXiv:0704.2457 [hep-ph]].
\bibitem{Stephanov:2007ry}
M.~A.~Stephanov,
%``Deconstruction of Unparticles,''
Phys.\ Rev.\  D {\bf 76}, 035008 (2007) [arXiv:0705.3049 [hep-ph]].
\bibitem{CKY}
K.~Cheung, W.~Y.~Keung and T.~C.~Yuan,
  %``Novel signals in unparticle physics,''
 Phys. Rev. Lett. {\bf 99}, 051803 (2007) [arXiv:0704.2588 [hep-ph]]; arXiv:0706.3155
 [hep-ph].

\bibitem{LZ}
M.~X.~Luo and G.~H.~Zhu,
  %``Some Phenomenologies of Unparticle Physics,''
  arXiv:0704.3532 [hep-ph];
M.~X.~Luo, W.~Wu and G.~H.~Zhu,
  %``Unparticle Physics and A_{FB}^b on the Z pole,''
  arXiv:0708.0671 [hep-ph].

\bibitem{CH12}
 C.~H.~Chen and C.~Q.~Geng,
  %``Unparticle physics on CP violation,''
  arXiv:0705.0689 [hep-ph];
  Phys. Rev. D{\bf 76}, 036007 (2007)  [arXiv:0706.0850 [hep-ph].
\bibitem{chengeng}
  C.~H.~Chen and C.~Q.~Geng,
  %``Flavors and Phases in Unparticle Physics,''
  arXiv:0709.0235 [hep-ph].
\bibitem{Ding_Yan} G.~J.~Ding and M.~L.~Yan,
  %``Unparticle Physics in DIS,''
  arXiv:0705.0794 [hep-ph]; arXiv:0706.0325 [hep-ph].

\bibitem{Y_Liao}Y.~Liao,
  %``Bounds on unparticles couplings to electrons: From electron g-2 to
  %positronium decays,''
  arXiv:0705.0837 [hep-ph]; arXiv:0708.3327 [hep-ph];
% Long-ranged spin-spin interaction of electron from unparticle exchange
  Y.~Liao and J.~Y.~Liu, arXiv:0706.1284 [hep-ph].

\bibitem{Aliev}
  T.~M.~Aliev, A.~S.~Cornell and N.~Gaur,
  %``Lepton Flavour Violation in Unparticle Physics,''
  arXiv:0705.1326 [hep-ph]; JHEP {\bf 07}, 072 (2007)
   [arXiv:0705.4542 [hep-ph].

\bibitem{Catterall}
 S. Catterall and F. Sannino, Phys. Rev. D{\bf 76}, 034504 (2007)
 [arXiv:0705.1664 [hep-lat]].

\bibitem{Li}
  X.~Q.~Li and Z.~T.~Wei,
  %``Unparticle physics effects on D0 - anti-D0 mixing,''
  Phys. Lett. B{\bf 651}, 380 (2007) [arXiv:0705.1821 [hep-ph]];  arXiv:0707.2285
  [hep-ph].

\bibitem{Lu}
  C.~D.~Lu, W.~Wang and Y.~M.~Wang,
  %``Lepton flavor violating processes in unparticle physics,''
  arXiv:0705.2909 [hep-ph].

\bibitem{Fox}
  P.~J.~Fox, A.~Rajaraman and Y.~Shirman,
  %``Bounds on Unparticles from the Higgs Sector,''
  arXiv:0705.3092 [hep-ph].

\bibitem{Greiner}
  N.~Greiner,
  %``Constraints On Unparticle Physics In Electroweak Gauge Boson Scattering,''
  arXiv:0705.3518 [hep-ph].

\bibitem{Davou}
  H.~Davoudiasl,
  %``Constraining Unparticle Physics with Cosmology and Astrophysics,''
  arXiv:0705.3636 [hep-ph].

\bibitem{CGM}
   D.~Choudhury, D.~K.~Ghosh and Mamta,
  %``Unparticles and Muon Decay,''
  arXiv:0705.3637 [hep-ph].

\bibitem{XG}
  S.~L.~Chen and X.~G.~He,
  %``Interactions of Unparticles with Standard Model Particles,''
  arXiv:0705.3946 [hep-ph]; %  \bibitem{Chen:2007zy}
  S.~L.~Chen, X.~G.~He and H.~C.~Tsai,
  %``Constraints on Unparticle Interactions from Invisible Decays of Z,
  %Quarkonia and Neutrinos,''
  arXiv:0707.0187 [hep-ph].
  %%CITATION = ARXIV:0707.0187;%%


\bibitem{MR} P. Mathews and V. Ravindran,
  arXiv:0705.4599 [hep-ph].

\bibitem{Zhou}
S. Zhou, arXiv:0706.0302 [hep-ph].
% Title: Neutrino Decays and Neutrino Electron Elastic Scattering in
%Unparticle Physics

\bibitem{FFRS}
 %Minimal Walking Technicolor: Set Up for Collider Physics
 R. Foadi, M.~T. Frandsen, T.~A. Ryttov and F. Sannino, arXiv:0706.1696
 [hep-ph].

\bibitem{Bander}
  M.~Bander, J.~L.~Feng, A.~Rajaraman and Y.~Shirman,
  %``Unparticles: Scales and High Energy Probes,''
  arXiv:0706.2677 [hep-ph].

\bibitem{Rizzo}
  T.~G.~Rizzo,
  %``Contact Interactions and Resonance-Like Physics at Present and Future
  %Colliders from Unparticles,''
  arXiv:0706.3025 [hep-ph].

 \bibitem{GN} H.~Goldberg and P.~Nath,
  %``Ungravity and Its Possible Test,''
  arXiv:0706.3898 [hep-ph].

\bibitem{Zwicky}
  R.~Zwicky,
  %``Unparticles at heavy flavour scales: CP violating phenomena,''
  arXiv:0707.0677 [hep-ph].

\bibitem{Kikuchi}
  T.~Kikuchi and N.~Okada,
  %``Unparticle physics and Higgs phenomenology,''
  arXiv:0707.0893 [hep-ph].

\bibitem{M_Giri}
  R.~Mohanta and A.~K.~Giri,
  %``Unparticle effect on $B_s - \bar B_s$ mixing and its implications for $B_s
  %\to J/\psi \phi, \phi \phi$ decays,''
  arXiv:0707.1234 [hep-ph];
  arXiv:0707.3308 [hep-ph].

\bibitem{Huang}
  C.~S.~Huang and X.~H.~Wu,
  %``Direct CP violation of $B \to l \nu$ in unparticle physics,''
  arXiv:0707.1268 [hep-ph].

\bibitem{Krasnikov}
  N.~V.~Krasnikov,
  %``Unparticle as a field with continuously distributed mass,''
  arXiv:0707.1419 [hep-ph].

\bibitem{Lenz}
  A.~Lenz,
  %``Unparticle physics effects in B_s mixing,''
  arXiv:0707.1535 [hep-ph].

\bibitem{C_Ghosh} D.~Choudhury and D.~K.~Ghosh,
  %``Top off the unparticle,''
  arXiv:0707.2074 [hep-ph].

\bibitem{Zhang}
  H.~Zhang, C.~S.~Li and Z.~Li,
  %``Unparticle Physics and Supersymmetry Phenomenology,''
  arXiv:0707.2132 [hep-ph].

\bibitem{Nakayama}
  Y.~Nakayama,
  %``SUSY Unparticle and Conformal Sequestering,''
  arXiv:0707.2451 [hep-ph].

\bibitem{Deshpande} N.~G.~Deshpande, X.~G.~He and J.~Jiang,
  %``Supersymmetric Unparticle Effects on Higgs Boson Mass and Dark Matter,''
  arXiv:0707.2959 [hep-ph];  N.~G.~Deshpande, S.~D.~H.~Hsu and J.~Jiang,
  %``Long range forces and limits on unparticle interactions,''
  arXiv:0708.2735 [hep-ph].

\bibitem{DEQ}
  A.~Delgado, J.~R.~Espinosa and M.~Quiros,
  %``Unparticles-Higgs Interplay,''
  arXiv:0707.4309 [hep-ph].

\bibitem{Neubert}
  M.~Neubert,
  %``Unparticle Physics with Jets,''
  arXiv:0708.0036 [hep-ph].

\bibitem{Hannestad}
  S.~Hannestad, G.~Raffelt and Y.~Y.~Y.~Wong,
  %``Unparticle constraints from SN1987A,''
  arXiv:0708.1404 [hep-ph].

\bibitem{Das}
  P.~K.~Das,
  %``Unparticle effects in Supernovae cooling,''
  arXiv:0708.2812 [hep-ph].

\bibitem{Bhattacharyya}
  G.~Bhattacharyya, D.~Choudhury and D.~K.~Ghosh,
  %``Unraveling unparticles through violation of atomic parity and rare
  %beauty,''
  arXiv:0708.2835 [hep-ph].

\bibitem{Majumdar}
   D.~Majumdar,
  %``Unparticle decay of neutrinos and it's effect on ultra high energy
  %neutrinos,''
  arXiv:0708.3485 [hep-ph].

\bibitem{Alan-Pak}
  A.~T.~Alan and N.~K.~Pak,
  %``Unparticle pyhsics in top pair signals at the LHC,''
  arXiv:0708.3802 [hep-ph].

\bibitem{FD} A. Freitas and D. Wyler, arXiv:0708.4339 [hep-ph].
%Astro Unparticle Physics

\bibitem{Gogoladze:2007jn}
  I.~Gogoladze, N.~Okada and Q.~Shafi,
  %``Unparticle Physics And Gauge Coupling Unification,''
  arXiv:0708.4405 [hep-ph].

\bibitem{Hur:2007cr}
  T.~i.~Hur, P.~Ko and X.~H.~Wu,
  %``Antisymmetric rank--2 tensor unparticle physics,''
  arXiv:0709.0629 [hep-ph].

\bibitem{Anchordoqui:2007dp}
  L.~Anchordoqui and H.~Goldberg,
  %``Constraints on Unparticle Physics from Solar and KamLAND Neutrinos,''
  arXiv:0709.0678 [hep-ph].

\bibitem{Majhi:2007tu}
  S.~Majhi,
  %``W-pair production in Unparticle Physics,''
  arXiv:0709.1960 [hep-ph].

\bibitem{McDonald:2007bt}
  J.~McDonald,
  %``Cosmological Constraints on Unparticles,''
  arXiv:0709.2350 [hep-ph].

\bibitem{Kumar:2007af}
  M.~C.~Kumar, P.~Mathews, V.~Ravindran and A.~Tripathi,
  %``Diphoton production with Unparticle at LHC,''
  arXiv:0709.2478 [hep-ph].

\bibitem{Das:2007cc}
  S.~Das, S.~Mohanty and K.~Rao,
  %``Test of unparticle long range forces from perihelion precession of
  %Mercury,''
  arXiv:0709.2583 [hep-ph].

\bibitem{pdg}Particle Data Group, W.-M. Yao {\it et al.,} Journal of Physics G 33, 1
(2006)

\bibitem{Kuno:1999jp}
  Y.~Kuno and Y.~Okada,
  %``Muon decay and physics beyond the standard model,''
  Rev.\ Mod.\ Phys.\  {\bf 73}, 151 (2001)
  [arXiv:hep-ph/9909265].

\bibitem{Cheng:1976uq}
  T.~P.~Cheng and L.~F.~Li,
  %``Nonconservation Of Separate Mu - Lepton And E - Lepton Numbers In Gauge
  %Theories With V+A Currents,''
  Phys.\ Rev.\ Lett.\  {\bf 38}, 381 (1977);
  A.~Pilaftsis,
  %``Radiatively induced neutrino masses and large Higgs neutrino couplings in
  %the standard model with Majorana fields,''
  Z.\ Phys.\  C {\bf 55}, 275 (1992)
  [arXiv:hep-ph/9901206];
  A.~Masiero, S.~K.~Vempati and O.~Vives,
  %``Seesaw and lepton flavour violation in SUSY SO(10),''
  Nucl.\ Phys.\  B {\bf 649}, 189 (2003)
  [arXiv:hep-ph/0209303];
  J.~Hisano, T.~Moroi, K.~Tobe, M.~Yamaguchi and T.~Yanagida,
  %``Lepton flavor violation in the supersymmetric standard model with seesaw
  %induced neutrino masses,''
  Phys.\ Lett.\  B {\bf 357}, 579 (1995)
  [arXiv:hep-ph/9501407];
\bibitem{Ellis}J.~R.~Ellis, J.~Hisano, M.~Raidal and Y.~Shimizu,
  %``A new parametrization of the seesaw mechanism and applications in
  %supersymmetric models,''
  Phys.\ Rev.\  D {\bf 66}, 115013 (2002)
  [arXiv:hep-ph/0206110];A.~Brignole and A.~Rossi,
  %``Anatomy and phenomenology of mu tau lepton flavour violation in the
  %MSSM,''
  Nucl.\ Phys.\  B {\bf 701}, 3 (2004)
  [arXiv:hep-ph/0404211];E.~Arganda and M.~J.~Herrero,
  %``Testing supersymmetry with lepton flavor violating tau and mu decays,''
  Phys.\ Rev.\  D {\bf 73}, 055003 (2006)
  [arXiv:hep-ph/0510405].

\bibitem{Nardi:1992nq}
  E.~Nardi,
  %``Z-prime, new fermions and flavor changing processes. Constraints on E(6)
  %models from mu $\to$ e e e,''
  Phys.\ Rev.\  D {\bf 48}, 1240 (1993)
  [arXiv:hep-ph/9209223];B.~Murakami,
  %``The impact of lepton-flavor violating Z' bosons on muon g-2 and  other muon
  %observables,''
  Phys.\ Rev.\  D {\bf 65}, 055003 (2002)
  [arXiv:hep-ph/0110095];
  J.~Bernabeu, E.~Nardi and D.~Tommasini,
  %``mu - e conversion in nuclei and Z-prime physics,''
  Nucl.\ Phys.\  B {\bf 409}, 69 (1993)
  [arXiv:hep-ph/9306251].

%\bibitem{Aubert:2005ye}
%  B.~Aubert {\it et al.}  [BABAR Collaboration],
%  %``Search for lepton flavor violation in the decay $\tau \to \mu \gamma$,''
%  Phys.\ Rev.\ Lett.\  {\bf 95}, 041802 (2005)
%  [arXiv:hep-ex/0502032].

\bibitem{Dohmen:1993mp}
  C.~Dohmen {\it et al.}  [SINDRUM II Collaboration.],
  %``Test Of Lepton Flavor Conservation In Mu $\to$ E Conversion On Titanium,''
  Phys.\ Lett.\  B {\bf 317}, 631 (1993).
\bibitem{Bertl:2001fu}
  W.~Bertl {\it et al.},
  %``Search for muon - electron conversion on gold,''
  Eur.\ Phys.\ J.\  C {\bf 47} (2006) 337.

\bibitem{Ritt:2006cg}
  S.~Ritt  [MEG Collaboration],
  %``Status Of The Meg Expriment Mu $\to$ E Gamma,''
  Nucl.\ Phys.\ Proc.\ Suppl.\  {\bf 162}, 279 (2006).

\bibitem{mueconv}Y. Mori et al. [PRISM/PRIME working group], LOI at J-PARC 50-GeV PS, LOI-
25, http://psux1.kek.jp/~jhf-np/LOIlist/LOIlist.html.

\bibitem{Kitano:2002mt}
R.~Kitano, M.~Koike and Y.~Okada,
%``Detailed calculation of lepton flavor violating muon electron  conversion
%rate for various nuclei,''
Phys.\ Rev.\  D {\bf 66}, 096002 (2002) [arXiv:hep-ph/0203110].

\bibitem{Czarnecki:1998iz}
  A.~Czarnecki, W.~J.~Marciano and K.~Melnikov,
  %``Coherent muon electron conversion in muonic atoms,''
  AIP Conf.\ Proc.\  {\bf 435}, 409 (1998)
  [arXiv:hep-ph/9801218].

\bibitem{Kosmas:1990tc}
  T.~S.~Kosmas and J.~D.~Vergados,
  %``Study Of The Flavor Violating (Mu-, E-) Conversion In Nuclei,''
  Nucl.\ Phys.\  A {\bf 510}, 641 (1990);
  H.~C.~Chiang, E.~Oset, T.~S.~Kosmas, A.~Faessler and J.~D.~Vergados,
  %``Coherent And Incoherent (Mu-, E-) Conversion In Nuclei,''
  Nucl.\ Phys.\  A {\bf 559}, 526 (1993).
\bibitem{ding} Gui-Jun Ding and Mu-Lin Yan, in progress.


\end{thebibliography}
\end{document}